\documentclass[twocolumn,showpacs,preprintnumbers,superscriptaddress,amsmath,amssymb]{revtex4}

\usepackage{graphicx}
\usepackage{dcolumn,color}
\usepackage{bm}
\usepackage{colortbl}

\begin{document}

\preprint{APS/123-QED}

\title{Electromagnons in the multiferroic state of perovskite manganites with symmetric-exchange striction}

\author{Y. Takahashi} 
 \affiliation{Multiferroics Project, ERATO, Japan Science and Technology Agency (JST)
 c/o Department of Applied Physics, The University of Tokyo, Tokyo 113-8656, Japan} 
\author{S. Ishiwata}
 \affiliation{Cross-correlated Materials Research Group (CMRG), ASI, RIKEN, Wako 351-0198, Japan} 
\author{S. Miyahara}
 \affiliation{Multiferroics Project, ERATO, Japan Science and Technology Agency (JST)
 c/o Department of Applied Physics, The University of Tokyo, Tokyo 113-8656, Japan}
\author{Y. Kaneko}
 \affiliation{Multiferroics Project, ERATO, Japan Science and Technology Agency (JST)
 c/o Department of Applied Physics, The University of Tokyo, Tokyo 113-8656, Japan}
\author{N. Furukawa}
  \affiliation{Multiferroics Project, ERATO, Japan Science and Technology Agency (JST)
 c/o Department of Applied Physics, The University of Tokyo, Tokyo 113-8656, Japan}
  \affiliation{Department of Physics and Mathematics, Aoyama Gakuin University, Sagamihara, Kanagawa 229-8558, Japan}
\author{Y. Taguchi}
\affiliation{Cross-correlated Materials Research Group (CMRG), ASI, RIKEN, Wako 351-0198, Japan} 
\author{R. Shimano}
 \affiliation{Multiferroics Project, ERATO, Japan Science and Technology Agency (JST)
 c/o Department of Applied Physics, The University of Tokyo, Tokyo 113-8656, Japan}   
 \affiliation{Department of Physics, The University of Tokyo, Tokyo 113-0033, Japan} 
\author{Y. Tokura}
 \affiliation{Multiferroics Project, ERATO, Japan Science and Technology Agency (JST)
 c/o Department of Applied Physics, The University of Tokyo, Tokyo 113-8656, Japan} 
 \affiliation{Cross-correlated Materials Research Group (CMRG), ASI, RIKEN, Wako 351-0198, Japan}
 \affiliation{Department of Applied Physics, The University of Tokyo, Tokyo 113-8656, Japan}

\date{\today}

\begin{abstract}
We have investigated electrically-active magnetic excitations (electromagnons) in perovskite manganites with the $E$-type (up-up-down-down) spin structure by terahertz spectroscopy. Eu$_{1-x}$Y$_x$MnO$_3$ (0.1$\le x\le$1) and Y$_{1-y}$Lu$_y$MnO$_3$ (0$\le y\le$1) without magnetic $f$-moments, which host collinear sinusoidal, $A$-type, cycloidal, and $E$-type spin orders, are used to examine the systematics of possible electromagnons. Three-peak structures (23, 35, 45 cm$^{-1}$) of magnetic origin show up in the $E$-type phase with little composition ($y$) dependence of frequencies, making a contrast with the electromagnons observed in the  cycloidal-spin ($x\le0.8$) phases. One of these electromagnon is ascribed to the zone-edge magnon mode based on the calculated magnon dispersions.
\end{abstract}

\maketitle

Renewed attention to multiferroics, 
where the electric and magnetic orders can coexist, 
has been aroused since the discovery of the electric-polarization flop 
by applying magnetic field in a perovskite manganite \cite{Kimura}. The series of perovskite $R$MnO$_3$ ($R$=Tb, Dy, Eu$_{1-x}$Y$_x$) have a helical (cycloidal) structure of Mn$^{3+}$ spins due to the frustration between ferromagnetic (nearest-neighbor) $J_1$ and antiferromagnetic (second nearest-neighbor) $J_2$ exchange interactions on the $ab$ plane. The magnetically induced ferroelectricity has been explained by the inverse Dzyaloshinski-Moriya (IDM) interaction \cite{Katsura1, Mostovoy, Sergienko}, in which the microscopic polarization $\bm {P}$ is expressed as $\bm {P}\propto \bm {e}_{ij}\times(\bm {S}_i\times \bm {S}_j)$ for a pair of neighboring spins $\bm {S}_i$, $\bm {S}_j$ with the unit vector  $\bm {e}_{ij}$ connecting them. The cycloidal spins accordingly produce the macroscopic (spontaneous) polarization ($P_s$) parallel to the spiral plane and perpendicular to the magnetic modulation vector $q_m$. In contrast, perovskite manganites $R$MnO$_3$ with smaller rare-earth ionic radius ($R$=Ho, Tm, Yb, Lu, Y$_{1-y}$Lu$_y$) exhibit a commensurate $E$-type spin structure: up-up-down-down in the $ab$ plane \cite{Kimura2}, hosting also the ferroelectricity. The ferroelectricity in the $E$-type phase originates from the symmetric-exchange term expressed as $\bm {S}_i\cdot \bm {S}_j$, which potentially produces larger $P_s$ than IDM mechanism \cite{Picozzi}. In fact, a $P_s$ of 4 times as large as that of the $ab$-plane cycloidal-spin phase has been reported for the $E$-type phase \cite{Pomjakushin, Ishiwata}. Thus, a whole series of perovskite can show versatile multiferroic states, where the two representative but different mechanisms --symmetric and antisymmetric exchange strictions-- are effective. 

The dynamical interplay between magnetic and electric excitations has also been expected for multiferroic materials. The first observation of electric-dipole active magnetic excitation, termed ''$electromagnon$ (EM)'', was reported for TbMnO$_3$ as far-infrared absorption around 20 cm$^{-1}$ \cite{Pimenov1}. Successive investigations have clarified that the strong infrared absorption showing up only for $\bm {E}^{\omega} \| a$ with a two-peak structure is common to the EM in the cycloidal-spin phases of $R$MnO$_3$ irrespective of the $ab$- or $bc$-spiral plane \cite{Valdes1, Kida1, Pimenov2, Kida2, Takahashi1, Takahashi2, Pimenov3}.  The initially predicted EM, which corresponds to a rotary oscillation mode of the spiral-spin plane ({\it i.e.} a vibration of $P_s$), should show the polarization selection rule depending on the direction of $P_s$ ($\bm {E}^\omega \| c$ for $\bm {P}_s \| a$, or $\bm {E}^\omega \| a$ for $\bm {P}_s \| c$)\cite{Katsura2, Cano}; however, this is in contradiction with the experimental observation \cite{Kida1}. To settle the discrepancy, the symmetric exchange striction (SES) mechanism has been proposed for the EM activity \cite{Valdes2, Miyahara}; a noncollinear spin structure like the cycloid gives rise to the electric dipole activity ($\bm {E}^\omega \| a$) of the zone-edge magnon mode via the local electric dipole $\bm {\pi}_{ij}(\bm {S}_i\cdot \bm {S}_j)$ built-in the $i$- and $j$-th Mn sites. 
In fact, the energy of the higher-lying mode at 60 cm$^{-1}$ in TbMnO$_3$ is consistent with 
the zone-edge mode energy in the magnon dispersion 
revealed by inelastic neutron scattering \cite{Senff, Miyahara, Valdes2}. 
The observed feature that the higher-lying EM-peak energy lowers with decreasing the radius of A-site ion shows a good agreement with the calculated result based on the Heisenberg model \cite{Lee}. 
Recently, the model for lower-lying mode was suggested based also on the SES mechanism \cite{Stenberg}.
Since the collinear spin structure cannot give rise to any EM in the SES mechanism, the suppression of the EM in the $A$-type phase \cite{Pimenov2, Kida2, Takahashi2} can be reasonably explained by this mechanism. As far as the SES mechanism is concerned, no electric activity is expected for any one-magnon mode for the {\it genuine}, {\it i.e.} perfectly collinear, $E$-type phase.

\begin{figure}[!htb]
\includegraphics[ width = 1 \linewidth]{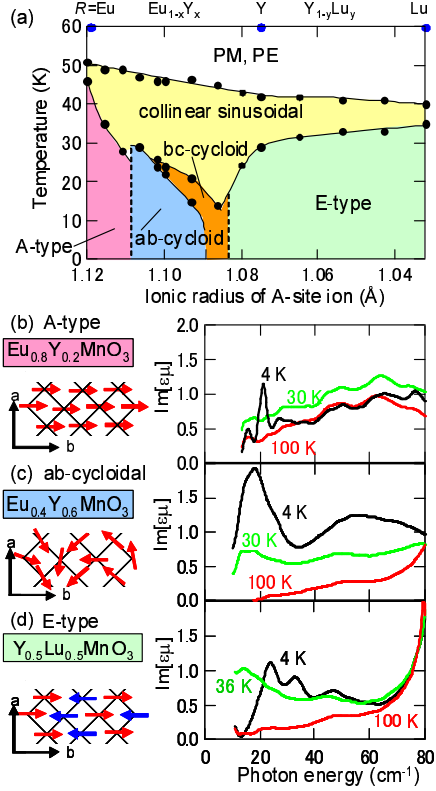}
\caption{\label{fig:epsart} (color online) (a) The phase diagram of Eu$_{1-x}$Y$_x$MnO$_3$ (0.1$\le x\le$1) and Y$_{1-y}$Lu$_y$MnO$_3$ (0$\le y\le$1) reproduced from Refs.[\onlinecite{Hemberger, Yamasaki, Ishiwata}]. The Im[$\epsilon\mu$] spectra for (b) $A$-type, (c) $ab$-cycloidal and (d) $E$-type spin phase with the schematic spin structures.}
\end{figure} 

In this Letter, we have systematically investigated the EM in the $E$-type spin phase on the basis of the comprehensive phase diagram established for Eu$_{1-x}$Y$_x$MnO$_3$  (0.1$\le x\le$1) and Y$_{1-y}$Lu$_y$MnO$_3$ (0$\le y\le$1) with no $f$-electron magnetic moment on the perovskite A-site \cite{Ishiwata}, by using terahertz time-domain spectroscopy (THz-TDS). In these materials, we can avoid the effect of the crystal field excitations of rare-earth ions, such as $R=$Ho, Er, Tm and Yb to unambiguously confirm the existence of the well-defined EMs in the $E$-type phase.

The high-quality polycrystalline samples were synthesized under high pressure; the details for synthesis are described elsewhere \cite{Ishiwata}. Specularly polished samples with 3 mm in diameter and 500 $\mu$m in thickness were prepared. The terahertz wave was generated by the optical rectification of an ultrashort laser pulse with a ZnTe(110) crystal and detected by a dipole antenna \cite{Ferguson}. From the measured complex transmittance spectrum at 13 - 80 cm$^{-1}$, we determined the complex optical constant $n^2=\epsilon\mu$, where the $\epsilon$ and $\mu$ are complex dielectric and magnetic permittivity, respectively. In the analysis, we assume that $\mu \approx 1$ and its negligible effect on the Fresnel reflection coefficient. The validity of this assumption was checked in Ref. \cite{Kida1}, where the detailed experimental setup was also described. The possible spectral modification was carefully checked by comparing the spectra of the both single-crystalline and polycrystalline samples of DyMnO$_3$ and Eu$_{0.8}$Y$_{0.2}$MnO$_3$.

Figure 1(a) shows the phase diagram of Eu$_{1-x}$Y$_x$MnO$_3$ and Y$_{1-y}$Lu$_y$MnO$_3$ system \cite{Hemberger, Yamasaki, Ishiwata}. The representative spectra of the $A$-type ($x$=0.2), $ab$-cycloidal ($x$=0.6) and $E$-type ($y$=0.5) states are displayed in Fig. 1(b, c, d) with the respective schematic spin configurations. The plateau-like absorption bands are observed in the higher-T collinear sinusoidal phase for all the compounds (30 K for $x$=0.2, 0.6, and 36 K for $y$=0.5) in addition to the tail of the higher-lying phonon modes; this broadened band has been assigned to the precursory EM-like excitation arising from the cycloidal spin fluctuation \cite{Takahashi2}. The suppression of this excitation band is seen upon the transition to the $A$-type phase (4 K) for $x$=0.2 (Fig. 1(b)). The sharp antiferromagnetic resonance (AFMR) with much smaller spectral intensity, which is active for the magnetic field component of light, is discerned at 20 cm$^{-1}$ in the $A$-type phase \cite{Pimenov2, Kida2, Takahashi2}. In contrast, the development of two clear peaks (18 and 54 cm$^{-1}$) is identified in the cycloidal-spin phase (4 K) for $x$=0.6 (Fig. 1 (c)), similarly to the EM spectra for the cycloidal-spin phases of other $R$MnO$_3$ \cite{Lee}. For $y$=0.5 with the $E$-type phase, a new EM-like feature is revealed in this study (Fig. 1(d)); three pronounced peaks appear at 24, 33 and 48 cm$^{-1}$, that are in a frequency region typical of magnetic (magnon) excitations. A relatively large spectral weight as developed below 100 K suggests their EM nature.

 \begin{figure}
\includegraphics[ width =1 \linewidth]{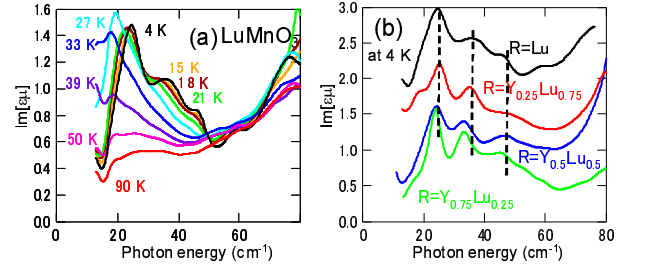}
\caption{\label{fig:epsart} (color online) (a) Temperature dependence of Im[$\epsilon \mu$] spectra for LuMnO$_3$. (b) Im[$\epsilon \mu$] spectra of the $E$-type phase at 4 K for Y$_{1-y}$Lu$_y$MnO$_3$ ($y$=0.25, 0.5, 0.75, and 1) with vertical offset for clarity. Peak positions are indicated by the vertical dotted lines.}
\end{figure} 

 Figure 2(a) shows the temperature dependence of Im[$\epsilon\mu$] spectra for LuMnO$_3$ ($y$=1). At 90 K (paramagnetic phase), only the contribution from the low energy tail of the higher-lying phonon modes can be seen. The broad absorption band develops from slightly-above $T_N$=40 K with decreasing temperature, and then it shows rapid enhancement below $T_N$. This material undergoes the transition from the collinear sinusoidal to the $E$-type phase at $T_c$= 35 K with lock of $q_m$ from incommensurate (IC) to commensurate, accompanied by the emergence of ferroelectricity. In accordance with this phase transition, the broad absorption band turns into peak structures at 24, 38 and 46 cm$^{-1}$ with slight hardening of the frequencies while decreasing temperature. The spectra at the lowest temperature (4 K) for the $E$-type phase are displayed for all the measured samples (Y$_{1-y}$Lu$_y$MnO$_3$) in Fig. 2(b). All the spectra show the three-peak feature in common, being distinct from the feature of the two broad peaks in the IC cycloidal-spin phase (Fig. 1(c)).

\begin{figure}
\includegraphics[ width = 1 \linewidth]{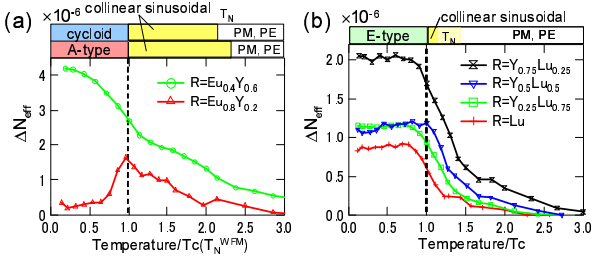}
\caption{\label{fig:epsart} (color online) Temperature dependence of spectral weight ($\Delta N_{\rm eff}$; see text for definition) of the electromagnon (or electric-dipole active magnetic excitation); (a) $A$-type ($x$=0.2), $ab$-cycloidal ($x$=0.6) for Eu$_{1-x}$Y$_x$MnO$_3$, and (b) $E$-type spin phase, $y$=0.25, 0.5, 0.75, and 1 for Y$_{1-y}$Lu$_y$MnO$_3$.}
\end{figure}

For more quantitative discussion, we evaluate the integrated spectral weight per Mn site, as given by
\begin{equation}
N_{\rm eff}=\frac{2m_0V}{\pi e^2}\int ^{\omega2}_{\omega1}\omega 'Im[\epsilon\mu (\omega ')] d\omega ',
\end{equation}                          
where $m_0$, $e$ and $V$ are the bare electron mass, charge and the unit-cell volume, respectively. Spectral weight of the EM or electric-dipole active magnetic excitations \cite{EDAME} ($\Delta N_{\rm eff}$) is defined as the increase of $N_{\rm eff}$ (between $\omega _1=13$ and $\omega _2$=70 cm$^{-1}$) from 100 K so as to eliminate the non-magnetic contributions. Figure 3(a) shows the dependence of $\Delta N_{\rm eff}$ on the temperature normalized by $T_c$=21 K for $x$=0.6 and by $T^{WFM}_N$=28 K for $x$=0.2 in Eu$_{1-x}$Y$_x$MnO$_3$. After the gradual increase of $\Delta N_{\rm eff}$ in the collinear sinusoidal phase for the both compounds, a rapid increase is observed for $x=0.6$, while a clear suppression for $x=0.2$ occurs, below the respective transition temperatures. (A small value of $\Delta N_{\rm eff}$ ($\sim 0.1\times 10^{-6}$) for AFMR is discerned in the $A$-type phase for $x$=0.2.) In contrast, $\Delta N_{\rm eff}$ in the $E$-type phase below $T_c$ keeps a nearly constant value for all the compositions (see Fig. 3(b)), showing again an essential difference from the EM in the IC cycloidal-spin phase. The electric activity of absorption bands in the $E$-type phase is again manifested by a fairly large $\Delta N_{\rm eff}$ compared to that of AFMR.

\begin{figure}
\includegraphics[ width = 1 \linewidth]{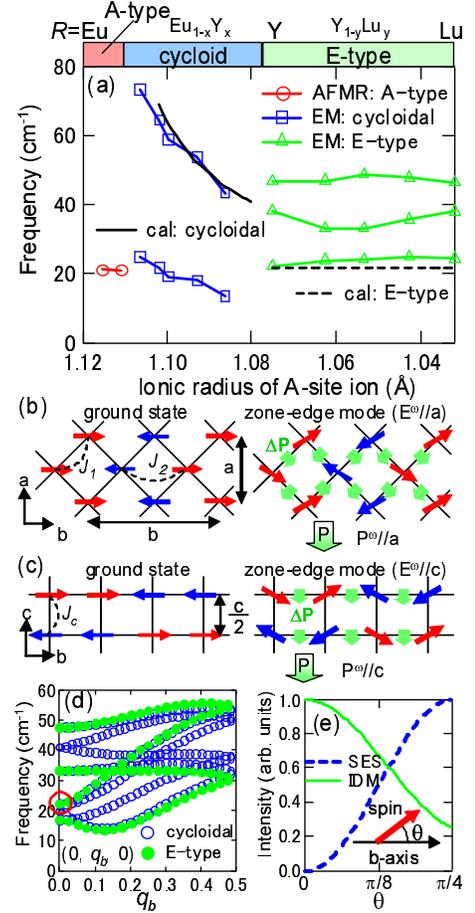}
\caption{\label{fig:epsart} (color online) (a) The composition ($x$ and $y$) dependence of frequencies of EM for the cycloidal and $E$-type spin phases (or of AFMR for $A$-type spin phase). The energies of EM obtained by calculation are also included. 
The schematic spin configuration of the $E$-type state and zone-edge magnon in (b) $ab$-plane and (c) $bc$-plane. The magnetic parameters $J_1$, $J_2$ and $J_c$, and unit cell length along $a$, $b$, and $c$ axes are displayed. 
(d) The magnon dispersions calculated for the $E$-type ($\theta = 0$)
and cycloidal phase ($\theta = \pi/4$) in the case of the A-site ionic radius of 1.05 $\AA$ (see text). The Brillouin zone along $q_b$ is a half of that for the chemical lattice (see Fig. 4(b)), and hence the zone-edge mode is folded into $q_b$=0.
The EM is indicated by an open circle (red online).
(e) Tilting angle $\theta$ dependence of the normalized absorption intensity 
by SES and IDM mechanisms.}
\end{figure} 

The composition ($x$ and $y$ in Eu$_{1-x}$Y$_x$MnO$_3$ and Y$_{1-y}$Lu$_y$MnO$_3$) dependence of the frequencies of EM and AFMR is plotted in Fig. 4(a). In the cycloidal phase, the frequency of the higher-lying mode strongly depends on the compositions, reflecting the sensitivity of the zone-edge magnon energy to the A-site ionic radius ($r_A$) or equivalently to the spin exchange energy \cite{Lee}. 
In fact, we can reproduce the systematic change in the frequency of 
the zone-edge EM mode in the light of the SES mechanism by assuming the exchange interactions 
as a function of $r_A$: $J_{\alpha} = c_{1\alpha} + c_{2\alpha} {r_A}$ ($\alpha = 1,2$,and $c$) 
in a 3-dimensional Heisenberg model (see Fig. 4 (b, c) for definition of $J_{\alpha}$; $c_{1\alpha}$ and $c_{2\alpha}$ are constants) \cite{Lee}.
In contrast, the frequencies of all EM modes in the $E$-type phase show little dependence on $r_A$ despite the systematic lattice-structural variation \cite{Ishiwata}. 
In this phase, we propose two possible mechanisms to explain the electric activity of zone-edge magnon based on the symmetric-exchange and antisymmetric-exchange interactions. In the $E$-type phase, the collinear spin structure parallel to the $b$ axis was proposed on the basis of the neutron scattering for R=Ho, Tm, Yb and Lu \cite{Pomjakushin, Neutron}. 
However, single ion anisotropy or DM interaction may be effective to tilt the spin directions off the $b$ axis \cite{Mochizuki}, resulting in the noncollinear spin state as a highly ellipsoidal cycloid. In such a case, the EM can show up for $\bm {E}^\omega \| a$ via the SES mechanism as has been observed in the IC cycloidal-spin phase. 
On the other hand, the antisymmetric-exchange interaction can induce the EM in the collinear $E$-type state. 
For example, the zone-edge magnon mode in the collinear $E$-type ground state turns out to take the dynamical cycloid like spin configuration on the $ab$ plane, which can induce uniform but dynamical polarization $P^{\omega}\| a$ via the IDM interaction, as shown in Fig. 4(b). Similarly, the zone-edge mode in the $bc$ plane induces $\bm {P}^\omega \| c$, as shown in Fig. 4(c). Although the electric activity of these modes ($\bm {P}^\omega \| a$ and $\bm {P}^\omega \| c$) would be distinguished by the polarization selection rule ($\bm {E}^\omega \| a$ and $\bm {E}^\omega \| c$, respectively) in a single crystal, while it is not possible at the moment because of the polycrystalline nature of the samples.

To get further insight into the nature of EM, we calculate the magnon dispersion for both the collinear $E$-type ($\theta=0$) and the commensurate cycloidal-spin state ($\theta=\pi/4$) (see inset of Fig. 4(e) for definition of $\theta$). 
As a typical case, we employed parameters of  $|J_1| = 5.6\,{\rm cm}^{-1}$, $J_2 = 15\,{\rm cm}^{-1}$, $J_c = 5.3\,{\rm cm}^{-1}$ and $D = 2.2\,{\rm cm}^{-1}$ for uniaxial anisotropy $D \sum (S_i^\alpha)^2$ for both ground states; the results are shown in Fig. 4(d).
The zone-edge mode has the same frequency ($\sim$23 cm$^{-1}$) for the both ground states as indicated by a circle in Fig. 4(d), resulting in the insensitive frequency to $\theta$.
Thus the observed lower-lying EM at 23 cm$^{-1}$ can be attributed to the zone-edge mode for any $\theta$.
Here, the decrease of $r_A$ enhances $J_2$, but hardly affects the frequency of zone-edge mode, which is consistent with the observed $r_A$ dependence of the EM frequency (Fig. 4(a)).
The $\theta$ dependence of absorption intensity is shown for the respective mechanisms in Fig. 4(e).
Thus, there are two possible scenarios as the origin of EM: (i) The IDM mechanism is dominant when $\theta \sim 0$. (ii) The SES mechanism, which likely gives a larger electric-dipole activity than the IDM mechanism, is dominant even for a small tilted angle $\theta$ (pseudo-$E$-type state).
The magnetic moment determined by neutron diffraction study for LuMnO$_3$ is 3.37 $\mu_B$/Mn alog $b$ axis \cite{Neutron}, implying the residual noncollinear component.
Although the twin EM model \cite{Stenberg} can be a candidate to induce a low-lying EM mode, too large anisotropy for $J_1$ with respect to the spin direction necessary for this model might be unrealistic.
The origin of two observed higher-lying modes around 35 and 46 cm$^{-1}$ would be assigned to the higher-lying $q_b$=0 modes if other magnetic terms were taken into account.

To conclude, we have investigated the electromagnon (EM) excitation in the $E$-type or pseudo-$E$-type (commensurate elliptical-spin) phase of the multiferroic perovskite manganite $R$MnO$_3$ with use of non-magnetic A-site ions systems, Eu$_{1-x}$Y$_x$MnO$_3$ (0.1$\le x\le$1) and Y$_{1-y}$Lu$_y$MnO$_3$ (0$\le y\le$1), by THz-TDS. The infrared absorption with three peaks (23, 35 and 45 cm$^{-1}$) is assigned to the EMs in the (pseudo-)$E$-type phase. The systematic investigation on the peak positions and spectral weight revealed the distinct nature of the EM in the $E$-type phase. We proposed two possibilities for the electric active zone-edge magnon in the $E$-type phase; the antisymmetric-exchange (IDM) and/or symmetric-exchange striction (SES) mechanism. The calculation of the magnon dispersion demonstrates the lower-lying zone-edge modes, whose energy coincides with the observed lowest lying mode.
 
The authors thank N. Nagaosa, T. Arima, N. Kida, M. Mochizuki and D. Okuyama for fruitful discussions.


\begin{thebibliography}{99}
\bibitem{EDAME} We use the term "electric-dipole active magnetic excitation" in a general sense, including the continuum-like excitation band.
\end{thebibliography}
\end{document}